\documentclass[12pt]{iopart}
\usepackage{xcolor}
\usepackage{csquotes}
\usepackage{float}

\begin{document}
\title[Impact of a primary school Einsteinian physics intervention]{ Long-term impact of a primary school intervention on aspects of Einsteinian physics}
\author{Kyla Adams$^1$, Roshan Dattatri, Tejinder Kaur$^1$, David Blair$^1$}
\address{$^1$ University of Western Australia, Crawley, WA 6009, Australia}
\ead{kyla.adams@uwa.edu.au}
\begin{abstract}
The physics that underpins modern technology is based on Einstein's theories of relativity and quantum mechanics. Most school students complete their compulsory science education without being taught any of these Einsteinian concepts. Only those who take a specialised physics course have the opportunity to learn modern physics.
In 2011, the first study of a modern physics teaching intervention with an Australian upper primary (aged 10--11) class was conducted. The initial intervention was the first step of the Einstein-First collaboration towards challenging the current paradigm of Newtonian teaching in schools. It was found that modern physics concepts could be taught to these students.
In 2020, 11 participants of the initial study (out of a total of 26) were contacted for a follow-up questionnaire and interview to investigate any long-term impact. The results of the follow-up indicate that the intervention maintained a positive impression on participants. The models and analogies used during the six-week intervention were highly memorable. The participants indicated that they found the intervention to be beneficial to their future learning. Even close to ten years after the intervention, the participants remembered several key concepts (such as curved space-time). The long-term follow-up indicates that Einsteinian physics can be taught at the upper primary level and be recalled several years later.
\end{abstract}
\submitto{\PED}
\maketitle
\section{Introduction}
%\wdcol{notes in red - things to add/fix and general questions}
Whether we are working at a computer or using GPS to find our way around a new city, the general theory of relativity and quantum physics impact our everyday lives. However, these concepts (collectively known as Einsteinian physics) are rarely included in compulsory science curricula \cite{ACARA2021,CDE2000,CBSE2017,So2014} despite their enormous relevance and significance.
\newline
\par\noindent
The public perception of Einsteinian physics is that it is challenging, difficult and abstract, especially without a rigorous mathematical background \cite{Foppoli2019,Angell2004}.
These perceptions have influenced physics teaching practices at all educational levels around the world. There is a growing interest in replacing the current teaching of the Newtonian paradigm with the modern Einsteinian paradigm \cite{Kersting2021}.\\

\par \noindent
To successfully achieve a paradigm shift, conflicting concepts between the old paradigm and the new have to be challenged, the relevance of the new concepts need to be addressed and the attitudes and perceptions of various stakeholders need to be challenged \cite{Treagust2018}. In the effort towards achieving a paradigm shift in physics education practices, the Einstein-First collaboration was created. The collaboration aims to prevent conceptual conflicts by teaching modern concepts from early in a students science education. By introducing a new Einsteinian paradigm students can be equipped with language, concepts and knowledge relevant to the modern world. For example, the recent  detection of gravitational waves from colliding black holes \cite{Abbott2016} cannot be meaningfully described within the Newtonian paradigm in which space is unable to curve. The Einstein-First project is based on the hypothesis that early introduction of Einsteinian physics is  beneficial because it prevents later contradictions, increases student motivation and fosters a positive attitude towards science. \\
\par \noindent
The first intervention conducted by the Einstein-First collaboration was at an Australian primary school. The intervention was presented to Year 6 students (aged 10--11) and consisted of six 45-minute lessons, a reading play, and an excursion to a Gravity Discovery Centre \footnote{More information about the Gravity Discovery Centre is available at https://gravitycentre.com.au}. These three aspects of the intervention encompass the Einstein-First aim of a simple and explicit introduction to Einsteinian physics concepts through hands-on and interactive activities \cite{Kaur2017a}. The interventions provide a conceptual basis for students that can be further built upon throughout their schooling. The reading play, entitled Free Float, was performed by the students at a local university. The initial results of this intervention can be found in \cite{Pitts2014}. \\
\par \noindent
The Einstein-First collaboration has since conducted multiple interventions in Australian schools with students from Year 3 (8--9 years old) to Year 10 (15--16) \cite{Baldy2007,Kaur2017c,Choudhary2018,Kaur2020}. Other short-term follow-up studies observed statistically significant positive influences in attitudes towards Einsteinian physics concepts \cite{Pitts2014, Kaur2017c}. A three-year longer-term follow-up found that there was a long-lasting effect on intervention participants \cite{Kaur2017d}. In addition, the Einstein-First methodology is suggested to have equalising effects in gender attitudes and understandings \cite{Kaur2020Gender}. Multiple interventions at schools in Australia and across the world \cite{Henriksen2014, Zahn2014} have been conducted with promising results, culminating in the creation of an international collaboration \cite{Choudhary2019} .
\\
\par \noindent
While these study results are very positive, they have only measured the short-term effects, with the exception of \cite{Kaur2017d} that studied Einsteinian concept retention of middle school students. Here we present a nine-year follow-up of a primary school  Einsteinian physics intervention. Our goal was to learn whether such early learning of Einsteinian concepts had significant student impact.
\\
\par \noindent
This paper presents results based on follow-up questionnaires and interviews with  participants, now aged 20--21, nine years after the first Einsteinian physics intervention in 2011. Through student networks we were able to contact about 40\% of the original class. While we acknowledge the small sample size, we believe that the results are interesting particularly as this was an isolated intervention and not supported by any follow-up learning.
\\
\par \noindent
The first section of this work presents the introduction to previous studies and the context for this follow-up study. \Sref{method} details the method used. \Sref{Results} reports on these findings in four subsections, \textit{Hands-on learning and reading play}, \textit{Effects on future learning}, \textit{Concepts remembered}, and \textit{Contradictions resulting from the intervention}. \Sref{Concl} is the \textit{Conclusion}.

\section{Method}\label{method}
The initial intervention focused on general relativity; the concept of space-time, the curvature of space, the warping of time and photons as particles that follow the shortest path in space-time. These concepts were investigated through three aspects, six in-class lessons, one excursion to a Gravity Discovery Centre and one excursion to a local university to perform a reading play \footnote{The play script is available at https://www.einsteinianphysics.com/wp-content/uploads/2020/06/Free-Float.pdf}. The intervention, presented in \cite{Pitts2014}, consisted of 26 participants. Of these 26 participants, 11 were able to be contacted for a discussion of their recollection of the intervention in a written response and an interview.
\subsection{The study}
In this study, a questionnaire was created by one of the researchers. Five open-ended questions were given to the participants of the initial intervention. The questions were:
\begin{enumerate}
    \item Please make a list of up to 10 things you remember from those courses.
    \item Did you learn things that prepared you for future learning at high school or university?
    \item Did you learn things that were later contradicted by teachers at school?
    \item Do you remember the \textit{play} [sic] you performed? Please tell us your impressions.
    \item Do you remember the excursions? What do you remember?
\end{enumerate}
The questionnaire was sent to one participant in the initial study. This participant was asked to pass it along to others in their network who were involved in the initial study. The responses were returned to a researcher, with some of these participants also agreeing to a follow-up phone interview. Another researcher conducted the interview. These interviews sought to elaborate on the five initial questions.

\subsection{Study sample}
The sample of participants consisted of eleven students from the initial intervention. The quotes and data presented here consist of the written responses (11) and recordings of the phone interviews (8). All participants who gave a phone interview also provided a written response. To protect participant anonymity, participants were randomly numbered from 1 to 11.

\subsection{Analysis}
The phone interviews were transcribed by a researcher and coded to match the five questions in the questionnaire. Participants were anonymised, and responses from the written and phone interview were collated following the methodology in \cite{Creswell2009}.

\subsection{Limitations}
In the ten years since the initial study, not all of the students could be contacted. The method of contacting the participants could have led to a selection bias in the samples, leading to only hearing from those who had positive recollections or had any memories. Further, while the sample selection was not strictly random, we could not identify any commonality between the participants. There are no common workplaces, universities, or other linking features that we could identify. This study is the first long-term (greater than three-years) follow-up by the Einstein-First collaboration and provides a starting point for the group to conduct rigorous long-term studies as the program rolls out to schools across Western Australia.

\section{Results}\label{Results}
The responses from the participants are grouped into four main themes that reflect the intentions of the initial study, they are \textit{Hands-on learning and reading play}, \textit{Effects on future learning}, \textit{Concepts remembered} and \textit{Contradictions resulting from the intervention}. All participant responses included at least one memory about the intervention, which is significant as the intervention occurred a little under ten years prior to being contacted for this study.
%-%-%-%-%-%-%-%-%-%-%-%-%-%-%-%-%-%-
\subsection{Hands-on learning and reading play}\label{play}

In the initial intervention the Einstein-First collaboration utilised interactive activities to challenge the preconception that Einsteinian physics is `difficult'. Various experiments, activities, and analogies were used, both in the classroom and during the excursion to the Gravity Discovery Centre (GDC) at Gingin, Western Australia \cite{Pitts2014}. This methodology is effective in changing preconceptions, which is necessary to achieve a paradigm shift \cite{Kaur2020,Kaur2018,Kaur2017a, Kaur2017b}. The participants were asked about these hands-on activities in the written questionnaire and follow-up phone interview.
\\
\par \noindent
All participants remembered at least one hands-on activity conducted during the intervention. Most of these remembered activities occurred at the GDC. All responses to question five about the excursion were positive. Participant 1 stated:
\begin{displayquote}
   `The fun experiments and activities I witnessed became fond memories as well as valuable lessons in how things such as gravity behave.' --Participant 1
\end{displayquote}
This quote is an example of how the hands-on learning, with models, analogies and activities, have shaped the perception of these traditionally `difficult' concepts and ideas. The positive memories of the hands-on activities are a step towards achieving a paradigm shift as they provide the context and relevance to the participants.
\\
\par\noindent
One of the most well-remembered activities was the experiment that consisted of dropping water balloons from a leaning tower at the GDC, with nine of the participants specifically mentioning it. Of those nine, two participants were also able to accurately recall the relevant concept. For example, in one response, when asked if they remembered the purpose of dropping water balloons from the leaning tower:
\begin{displayquote}
   `We dropped water balloons from the top of the tower to see if they would land at the same time.' --Participant 4
\end{displayquote}
\par\noindent
While most participants remembered dropping the water balloons, the experimental motivation was occasionally mis-remembered. One participant stated that the motivation for the drop tower was to demonstrate that space had curvature. While only present in one response, it does indicate potential confusion from the experiments. A larger sample size, across multiple interventions can aid in determining the sources for these confusions and correcting them. This is the aim of future studies.
\\
\par\noindent
Another activity mentioned by participants in relation to the GDC excursion was the lycra space-time simulator:
\begin{displayquote}
   `It was used to prove how mass can bend space-time ... and how different objects react' --Participant 10
\end{displayquote}
Participant 1 also discussed this activity:
\begin{displayquote}
   `I remember we had the golf-balls and you know when there was no golf-balls and you rolled the ball across it wouldn't really effect much but when there was a collection of balls in the middle it would you know, it would orbit essentially and you know the more mass there was, the steeper the curve you know, the more variation I guess.' --Participant 1
\end{displayquote}
In these examples participants could correctly relate the hands-on activity to the concept of curved space-time. These hands-on experiments allowed the participants to retain the knowledge of Einsteinian concepts.
\\
\par\noindent
All respondents indicated that they still have knowledge of the Einsteinian physics concepts from the intervention they covered when they were in Year 6. A participant summarised their memories of the hands-on learning approach in their comment:

\begin{displayquote}
   `Personally for me, a common theme was that actual experiments had a stronger impact on what I remember.' --Participant 8
\end{displayquote}
\noindent
A major hands-on component of the initial intervention was the reading play. The play explored the Einsteinian physics concepts of space, time, and gravity. In the process of performing the play, participants become familiar with these concepts and gain a sense of the historical context behind their development. The play also provides the participants with a novel learning opportunity outside of traditional classroom methods.
\\
\par \noindent
Of the five questions presented to the intervention participants, one of the questions relates to this reading play. In both the written and interview responses, the participant's recollections about the play were wide ranging, from remembering exact quotes, to having no recollection at all, even with further prompting. Table \ref{fig:Playdata} shows the overall participant responses. Five participants remembered their role or the role of a fellow student as well as a relevant concept. One participant could remember a concept from the play, but nothing else. Further, five participants had no recollection of the play - or vaguely remember that there was a play on prompting but could not recall any details.
\\
\begin{table}[H]
\caption{Comparison of participants recollection of their role and/or a concept of the reading play.}
\begin{indented}
    \item[]\begin{tabular}{@{}llll}
    \br%\hline
    Role + Concept & Concept Only & No Memory & Total \\
    \mr
     5 & 1 & 5 & 11 \\
    \br

\end{tabular}
\end{indented}
 \label{fig:Playdata}
 \end{table}

\par
\noindent
Generally, when a participant remembered the play they would also remember a relevant physical concept involved. Several unique Einsteinian physics concepts were recalled by the participants, from curved space-time to geometry. There was also a commonality between the participants being able to recollect what other participants said in the play, instead of what lines they had learnt. In their phone interview Participant 8 recalled:
\begin{displayquote}
    `I can't remember too much about the content, but I do remember the interaction so to speak between the characters for example - like it wasn't my part but I do kind of remember that like Einstein might be talking to someone else about some kind of concept and things like that.' --Participant 8
\end{displayquote}
The general attitude towards the play was amusement, fun and remembering concepts for the first time from these plays. Participant 1's written response summarises these attitudes,
\begin{displayquote}
    `I remember the play with fondness, and the message ``Ze space it is curved and ze time it is warped'' has remained in my head since, and since I started thinking about this idea several years before I formally studied it in high school, I believe it helped the concept ``click" in high school.' --Participant 1
\end{displayquote}
In their written responses, two participants said that they didn't remember the play; however, during the phone interview, when presented with leading questions about the play and the concepts, they were able to agree whether a concept was presented or not. These vague recollections indicate that the play may not have been as influential for some participants as others.
\\
\par
\noindent
Overall, the play reinforced some conceptual understanding for just over half of the participants (6 out of 11 students). The remainder (5 out of 11 students) could not recall the play at all or any specific details, even with prompting from a researcher. The hands-on activities and the reading play had a lasting impression on most participants. However, there were notable gaps across the sample. A larger sample size or more in-depth interview could help provide more detail and provide a more thorough representation. Generally, the hands-on learning approaches contribute to participants having a positive association with learning Einsteinian physics, even many years later.

%-%-%-%-%-%-%-%-%-%-%-%-%-%-%-%-%-%-%-%-
\subsection{Effects on future learning}\label{future}
Determining the effect of the intervention on participants future learning and attitude towards science is an important step towards achieving a paradigm shift.
Most participants (9 out of 11) stated that the intervention prepared them for future science learning. The general tone of responses was that the intervention helped place science learning into a positive light and made them open to learning more concepts in the future.
\\
\par\noindent
One participant refers to an example presented in the intervention, about the twin paradox and how it ended up being relevant in Year 12 physics. They say:
\begin{displayquote}
    `... I think that if I hadn't been introduced to these things about relativity...it would have taken me a lot longer to wrap my head around these concepts if I wasn't already introduced to them.' --Participant 1
\end{displayquote}
\par \noindent
None of the study participants did physics at a university level. One participant did physics up to a high school Year 12 level. Two of the participants who responded went onto a STEMM (Science, Technology, Engineering, Maths and Medicine) based degree or career. Overall, nine participants said that the intervention taught them things that prepared them for future learning at high school or university. Of these nine, many of them stated that the intervention provided them with a `good early basis and understanding of science' (Participant 11).
\\
\par\noindent
These responses indicate that the intervention shaped the participants attitudes towards science. However, two of the eleven participants said that they did not learn things that prepared them for high school and later. When asked to elaborate, these participants stated that they did not have a scientific background and didn't do anything related to physics or science at university, which is why they feel like it didn't prepare them. In future studies responses like these could be further discussed to determine any other effects on their learning.
\\
\par
\noindent
Overall, despite the intervention occurring just under ten years prior to the interviews being conducted, the participants still had positive associations with the intervention and overall believe that learning Einsteinian physics at 10--11 years old was beneficial to their learning experience. Further emphasised by another participant in their response:
\begin{displayquote}
    `The fact that I still remember the classes from almost 10 years ago (I don’t remember much else that I learnt in primary school) shows how beneficial they were to my learning experience.' --Participant 6
\end{displayquote}
\noindent
Another participant echoes this point in their response:
\begin{displayquote}
   `Yes, most of what we learned came up in our science classes in high school, which made it easier to remember and re learn.' --Participant 3
\end{displayquote}
\par
\noindent
These responses provide a positive reflection of the intervention and that learning Einsteinian physics early allows for new concepts to be presented and understood outside of a university context and makes Einsteinian physics accessible to more students.

%-%-%-%-%-%-%-%-%-%-%-%-%-%-%-%-%-%-
\subsection{Concepts remembered}\label{concept}

In the initial intervention \cite{Pitts2014} the authors examined the short-term impact by comparing pre- and post-test questionnaires that focused on the concepts presented during the intervention. In the long-term follow-up the participants were asked what concepts they remembered from the same intervention. All but one participant had specific answers to the question `Please make a list of up to 10 things you remember from these courses'. The one participant who didn't list specifics in their written response still recalled an activity when prompted during the phone interview. They remembered making liquid nitrogen ice-cream at one point during the intervention.
\\
\par\noindent
In response to this question, the majority of participants listed concepts related to space-time and curvature (16 mentions across the written and phone responses) (See \tref{fig:concepts}). The items listed ranged from the statement that space is curved to the equivalence principle. The next most mentioned concept was photons. In particular how they behave or their interaction with space-time. The concepts of photons and space-time are some of the most fundamental to Einsteinian physics and it is promising that they are the two most mentioned by the participants. While photons were not explicitly included in the initial intervention, the presenter would often mention that light is comprised of photons. Since the initial intervention photons have been explicitly included into the Einstein-First methodology. Including photons in the interventions had been shown to be effective in past, short-term Einstein-First interventions \cite{Kaur2017a}. The mention of photons by participants of the initial intervention further support the argument that this concept can be presented at the primary school level.\\
\par \noindent
Other concepts mentioned by the participants can be found in \tref{fig:concepts}. In addition, these concepts were unfamiliar to the participants at the beginning of the initial intervention, further emphasising the effect of the Einstein-First intervention. All of the concepts in \tref{fig:concepts} were explicitly covered in the intervention, except for magnetic force. However, during the excursion to the Gravity Discovery Centre the participants were exposed to multiple interactive physics activities, at least five of these related to magnetic force. \\
\par \noindent
Ten out of the eleven participants did not do physics at an upper high school level or beyond, the two most mentioned concepts (space-time and photons) are most likely a result of this intervention.

\begin{table}
    \caption{List of the different topics mentioned by the participants of the initial intervention with the number of total mentions across the written responses and phone interview. }
    \begin{indented}
    \item[]\begin{tabular}{@{} ll}
    \br
    Concept & Number of Mentions  \\
    \mr
    Space-time &	16 \\
    Photons & 7 \\
    Newton's Laws	& 2 \\
    Vacuum free fall	& 2 \\
    Black Holes	& 1 \\
    Acceleration and Velocity (differences) &	1 \\
    Distance and Displacement (differences) &	1 \\
    Magnetic force	& 1 \\
    Heisenberg Uncertainty Principle	& 1 \\
    $E = mc^2$ & 1\\
    Time is relative	& 1 \\
     Galileo and his experiments	& 1 \\
     \br
\end{tabular}
\end{indented}
    \label{fig:concepts}
\end{table}
\noindent
The broad spectrum of topics listed is a positive reflection of the combination of presentations, hands-on activities and excursions that were utilised during the intervention. This develops the argument that presenting Einsteinian physics concepts early provides a solid foundation that can easily be built upon. However, our sample size is limited, and future studies, with larger sample sizes, would aid in determining these effects.

%-%-%-%-%-%-%-%-%-%-%-%-%-%-%-%-
\subsection{Contradictions resulting from the intervention}\label{contradict}
One of the roadblocks to further introducing Einsteinian physics in the classroom is the preconception that the content will confuse participants, or that it will contradict what they will learn in the future. To investigate this, the question, `Did you learn things that were later contradicted by teachers at school?' was put to the participants. Overall most participants did not say that they learnt things that teachers later contradicted. Eight participants said that there were no contradictions. Of these eight, two participants were unable to further clarify why there were no contradictions. As a result these responses are limited.\\
\par \noindent
Two participants said that concepts presented in the intervention weren't contradicted later in their schooling, but they also didn't remember specifics. Responses to the other questions suggest these participants remembered the activities in a general sense, but, the specific concepts were less clear.
\\
\par\noindent
Of the eleven participants, one mentioned a contradiction between what was taught in this intervention and what they were taught later in school. Participant 4 stated that `the theory about the circumference of the circle' was contradicted in later years. In the phone interview, the same participant goes on to state:
\begin{displayquote}
   `There was an experiment to do with the circumference of a circle and we were shown that the, um, that the formula doesn't always apply when we applied the circumference around a balloon or something like that, but then obviously in high school we just got taught the standard formula.' --Participant 4
\end{displayquote}
The researcher conducting the interview further clarifies with the participant:
\begin{displayquote}
    Researcher: `So you were only taught the normal twice \textit{times} [sic] pi \textit{times} [sic] the radius in school whereas you were taught other stuff in this workshop?' \\
    Participant 4: `yeah yeah.'
\end{displayquote}
The recollection of the intervention's activity showed that it was significant and has embedded the understanding of curved space-time into this participant's idea. The later contradiction in their schooling is a result of Newtonian physics, and flat space-time being treated as the rule, not the exception. This confusion could be reduced by making it clear in the intervention that the `traditional' circumference calculation is correct, but only under specific circumstances.

\section{Conclusion}\label{Concl}
The first intervention conducted by the Einstein-First collaboration \cite{Pitts2014} found that there was a statistically significant improvement in student's conceptual understanding of Einsteinian physics, in the short-term \cite{Pitts2014} \footnote{Reference \cite{Pitts2014} found a mean increase of 2.0 marks (out of 20.0) with a 95\% confidence interval of 0.9-3.0 on student performance when comparing pre- and post-test results with a p value of 0.001 (from two-tailed paired-sample t test).}.\\
\par \noindent
This work followed-up on the participants of the initial intervention to determine if these significant effects were present in the long term. A limitation of the study is the number of participants, out of the initial 26 participants, 11 were able to be contacted. Selection bias of the respondents could be present. The participants had very positive memories of the intervention. Numerous participants could recall Einsteinian physics concepts many years later, despite not continuing physics study after their mandatory science education (Year 10 in Australia). Participants enjoyed the hands-on activities and found that they were useful in recalling the concepts. Half could recall the reading play, while the other half had no recollection, suggesting that it might not have been an effective teaching method. The sample size used in the study is quite low and may not be a full representation of the long term results. Despite this, there is a large consensus between responses across all the questions. The participants remembered several Einsteinian physics concepts and the related activities, years after the intervention. We conclude that Einsteinian physics can be taught at the upper primary level and be remembered many years later.
\\
\par\noindent

\ack
This research was supported by Australian Research Council Linkage Grant LP 180100859. One of us (KA) holds an “Australian Government Research Training Program Scholarship at The University of Western Australia”.
The authors would like to acknowledge the contribution of all the Einstein-First collaboration members, especially Carolyn Maxwell.
We also thank the students who participated in this program on a voluntary basis, both in the original study and this follow-up.
All participants voluntarily involved with this study gave their informed consent for the study and publication. This research was carried out under the University of Western Australia Ethics approval number 2019/RA/4/20/5875.

\newpage

\section*{References}
\bibliography{export}

\providecommand{\newblock}{}
\begin{thebibliography}{10}
\expandafter\ifx\csname url\endcsname\relax
  \def\url#1{{\tt #1}}\fi
\expandafter\ifx\csname urlprefix\endcsname\relax\def\urlprefix{URL }\fi
\providecommand{\eprint}[2][]{\url{#2}}

\bibitem{ACARA2021}
{Australian Curriculum, Assessment and Reporting Authority (ACARA)} 2010
  {Australian Curriculum, Assessment and Reporting Authority (ACARA)}
  \url{www.australiancurriculum.edu.au}

\bibitem{CDE2000}
{California Department of Education } 2000 {Science Content Standards for
  California Public Schools: Kindergarten through Grade 12}
  \url{www.cde.ca.gov/be/st/ss/documents/sciencestnd.pdf } date Acessed =
  22/02/2021

\bibitem{CBSE2017}
{Central Board of Education (CBSE)} 2017 {Indian Senior School Curriculum}
  \url{http://49.50.70.100/web\_material/Curriculum17/SrSecondary/07\%20Physics.pdf
  }

\bibitem{So2014}
So K and Kang K 2014 {\em Asia-Pacific Edu. Res.\/}

\bibitem{Foppoli2019}
Foppoli A, Choudhary R, Blair D, Kaur T, Moschilla J and Zadnik M 2018 {\em
  Physics Education\/} {\bf 54} 015001
  \urlprefix\url{https://doi.org/10.1088/1361-6552/aae4a4}

\bibitem{Angell2004}
Angell C, Øystein Guttersrud, Henriksen E~K and Isnes A 2004 {\em Science
  Education\/} {\bf 88}(5) 683--706 ISSN 00368326

\bibitem{Kersting2021}
Kersting M and Blair D 2021 {\em Teaching Einsteinian Physics in Schools\/}
  (Taylor and Francis (Routledge))

\bibitem{Treagust2018}
Treagust D, Won M and McLure F 2018 {\em Multiple Representations and
  Students’ Conceptual Change in Science\/} (London: Routledge) pp 121--128

\bibitem{Abbott2016}
Abbott B~P {\em et~al.\/} (LIGO Scientific Collaboration and Virgo
  Collaboration) 2016 {\em Phys. Rev. Lett.\/} {\bf 116}(6) 061102
  \urlprefix\url{https://link.aps.org/doi/10.1103/PhysRevLett.116.061102}

\bibitem{Kaur2017a}
Kaur T, Blair D, Moschilla J, Stannard W and Zadnik M 2017 {\em Physics
  Education\/} {\bf 52} 065012
  \urlprefix\url{https://doi.org/10.1088/1361-6552/aa83e4}

\bibitem{Pitts2014}
Pitts M, Venville G, Blair D and Zadnik M 2014 {\em Research in Science
  Education\/} {\bf 44}(3) 363--388 ISSN 15731898

\bibitem{Baldy2007}
Baldy E 2007 {\em International Journal of Science Education\/} {\bf 29}(14)
  1767--1788 ISSN 09500693

\bibitem{Kaur2017c}
Kaur T, Blair D, Moschilla J, Stannard W and Zadnik M 2017 {\em Physics
  Education\/} {\bf 52} 065014
  \urlprefix\url{https://doi.org/10.1088/1361-6552/aa83dd}

\bibitem{Choudhary2018}
Choudhary R~K, Foppoli A, Kaur T, Blair D~G, Zadnik M and Meagher R 2018 {\em
  Physics Education\/} {\bf 53} 065020
  \urlprefix\url{https://doi.org/10.1088/1361-6552/aae26a}

\bibitem{Kaur2020}
Kaur T, Blair D, Stannard W, Treagust D, Venville G, Zadnik M, Mathews W and
  Perks D 2020 {\em Research in Science Education\/} {\bf 50}(6) 2505--2532
  ISSN 15731898

\bibitem{Kaur2017d}
Kaur T, Blair D, Burman R, Stannard W, Treagust D, Venville G, Zadnik M,
  Mathews W and Perks D 2017 Evaluation of 14 to 15 year old students'
  understanding and attitude towards learning einsteinian physics
  (\textit{Preprint} \eprint{1712.02063})

\bibitem{Kaur2020Gender}
Kaur T, Blair D, Choudhary R, Dua Y, Foppoli A, Tragust D and Zadnik M 2020
  {\em Physics Education\/} {\bf 55}

\bibitem{Henriksen2014}
Henriksen E~K, Bungum B, Angell C, Tellefsen C~W, Fr{\aa}g{\aa}t T and B{\o}e
  M~V 2014 {\em Physics Education\/} {\bf 49} 678--684
  \urlprefix\url{https://doi.org/10.1088/0031-9120/49/6/678}

\bibitem{Zahn2014}
Zahn C and Kraus U 2014 {\em European Journal of Physics\/} {\bf 35}(5) ISSN
  13616404

\bibitem{Choudhary2019}
Choudhary R, Kraus U, Kersting M, Blair D, Zahn C, Zadnik M and Meagher R 2019
  {\em The Physics Educator\/} {\bf 01}(04) 1950016 ISSN 2661-3395

\bibitem{Creswell2009}
Creswell J 2009 {\em Research Design: Qualitative, Quantitative, and Mixed
  Methods Approaches\/} 3rd ed (SAGE)

\bibitem{Kaur2018}
Kaur T, Blair D, Stannard W, Treagust D, Venville G, Zadnik M, Mathews W and
  Perks D 2018 {\em Research in Science Education\/} {\bf 50}(6) 2505--2532
  ISSN 15731898

\bibitem{Kaur2017b}
Kaur T, Blair D, Moschilla J and Zadnik M 2017 {\em Physics Education\/} {\bf
  52} 065013 \urlprefix\url{https://doi.org/10.1088/1361-6552/aa83e1}

\end{thebibliography}

\end{document}